\documentclass[aps,prl,twocolumn]{revtex4-1}
\usepackage{graphicx}
\usepackage{amsmath}
\usepackage{dcolumn}
\usepackage{bm}
\usepackage{float}
\usepackage{color}

\usepackage{textcomp}

\makeatletter 
\renewcommand{\fnum@figure}{\textbf{Fig.~\thefigure}}
\makeatother

\begin{document}

\preprint{}

\title{Detection of thermodynamic ``valley noise'' in monolayer semiconductors: access to intrinsic valley relaxation timescales}

\author{M. Goryca$^{1,2}$, N. P. Wilson$^{3}$, P. Dey$^{1}$, X. Xu$^{3}$, S. A. Crooker$^{1}$}

\affiliation{$^1$National High Magnetic Field Laboratory, Los Alamos, NM 87545, USA}
\affiliation{$^2$Institute of Experimental Physics, Faculty of Physics, University of Warsaw, Warsaw, Poland}
\affiliation{$^3$Department of Physics, University of Washington, Seattle, WA 98195, USA}

\begin{abstract}

Together with charge and spin degrees of freedom, many new 2D materials also permit information to be encoded in an electron's valley degree of freedom -- that is, in particular momentum states in the material's Brillouin zone.  With a view towards future generations of valley-based (opto)electronic technologies, the intrinsic timescales of scattering and relaxation between valleys therefore represent fundamental parameters of interest.  Here we introduce and demonstrate an entirely passive, noise-based approach for exploring intrinsic valley dynamics in atomically-thin transition-metal dichalcogenide (TMD) semiconductors.  Exploiting the valley-specific optical selection rules in monolayer TMDs, we use optical Faraday rotation to detect, under conditions of strict thermal equilibrium, the stochastic thermodynamic fluctuations of the valley polarization in a Fermi sea of resident carriers. Frequency spectra of this spontaneous ``valley noise" reveal narrow Lorentzian lineshapes and therefore long exponentially-decaying intrinsic valley relaxation. Moreover, the valley noise signals are shown to validate both the relaxation times and the spectral dependence of conventional (perturbative) pump-probe measurements. These results provide a viable route toward quantitative measurements of intrinsic valley dynamics, free from any external perturbation, pumping, or excitation.

\end{abstract}

\maketitle

The development of novel two-dimensional materials such as graphene, bilayer graphene, and monolayer transition-metal dichalcogenide (TMD) semiconductors has rejuvenated long-standing interests \cite{Shkolnikov_2002_PRL_AlAs,Gunawan_2006_PRL} in harnessing valley degrees of freedom \cite{Xiao_2007_PRL_GrapheneValley,Rycerz_2007_NaturePhys_ValValveGraph,Culcer_2012_PRL,Zhu_2012_NaturePhys_bismuth,Isberg_2013_NatureMat_ValleytronicsDiamond,Xu_2014_NaturePhys_Review,Kormanyos_2015_2DMat_kpTheory,Schaibley_2016_NatureRevMat_Valleytronics}. Indeed, encoding information in an electron's momentum state (\textit{i.e.}, by which valley in the material's Brillouin zone the electron occupies) forms the conceptual basis of the burgeoning field of ``valleytronics''. In particular, the family of monolayer TMDs such as MoS$_2$ and WSe$_2$ have focused attention on valley physics because they provide a facile means of addressing specific valleys in momentum space using light: Owing to strong spin-orbit coupling and lack of crystalline symmetry, monolayer TMDs possess \textit{valley-specific} optical selection rules \cite{Yao_2008_PRB_GraphOptoelectronics,Xiao_2012_PRL,Cao_2012_NatureComm}, wherein right- and left-circularly polarized light couples selectively to transitions in the distinct $K$ and $K'$ valleys of their hexagonal Brillouin zone. Valley-polarized electrons, holes, and excitons can therefore be readily injected and detected optically, in marked contrast to most conventional semiconductors.

For any foreseeable valley-based information processing scheme, the intrinsic time scales of valley scattering and relaxation are of obvious critical importance. Recent optical pump-probe studies have shown that while valley-polarized electron-hole pairs (excitons) scatter very quickly on picosecond timescales \cite{Wang_2013_ACSNano_MoS2ValRelaxFast,Ochoa_2014_PRB_Theory,Wang_2014_PRB_WSe2ValRelaxFast,Yu_2014_PRB_Theory,Mai_2014_NanoLett_MoS2ValRelaxFast,Kumar_2014_Nanoscale_MoSe2ValRelaxFast,Singh_2016_PRL_WSe2TrionValRelaxFast}, the valley relaxation of \textit{resident} carriers in electron- or hole-doped TMD monolayers can be orders of magnitude longer \cite{Mak_2012_NatureNano_ControlValPolMoS2,Song_2013_PRL_TransportTMD,Hsu_2015_NatureComm,Yang_2015_NaturePhys,Song_2016_NanoLett,Volmer_2017_PRB,Yan_2017_PRB_WSe2Relax2ns,McCormick_2018_2Dmat}, even in the range of microseconds for holes at low temperatures \cite{Kim_2017_ScienceAdv_WSe2MoS2Relax40us,Dey_2017_PRL}. However, while very encouraging, all such pump-probe measurements are necessarily perturbative in nature. This is because optical pumping injects nonequilibrium electrons and holes which scatter, dissipate energy, and interact, thereby perturbing the resident carriers' valley polarization away from thermal equilibrium via processes not yet well understood.  Moreover, optical pumping of both majority \textit{and} minority carriers can create long-lived optically-inactive ``dark'' excitons and trions \cite{Molas_2017_2DMat_DarkX,Zhou_2017_NatureNano_DarkX,Zhang_2017_NatureNano_DarkX,Volmer_2017_PRB}, whose presence could in principle mask detection of carrier valley relaxation, as recently suggested \cite{Plechinger_2016_NatureComm,Volmer_2017_PRB}. An alternative means of accessing the truly \textit{intrinsic} valley relaxation of resident carriers in monolayer TMDs, free from perturbative effects, is therefore highly desired.

Fortunately, the fundamental relationship between a system's dynamic linear response (\textit{i.e.}, susceptibility) and its intrinsic fluctuations, as articulated by the fluctuation-dissipation theorem \cite{Kubo_1966_RPP}, suggests an alternative approach. Rather than perturb the valley polarization and measure its dissipative response, one could instead attempt to detect the \textit{spontaneous valley fluctuations} that necessarily must exist even in thermal equilibrium. If measurable, this thermodynamic ``valley noise'' will also encode the intrinsic relaxation timescales. Here we demonstrate that stochastic valley noise is indeed measurable in monolayer TMD semiconductors using optical Faraday rotation, and furthermore can be used as a powerful probe of the intrinsic valley dynamics in a Fermi sea of resident carriers, free from any external perturbation, pumping, or excitation.  

\begin{figure} [tbp]
\center
\includegraphics[width=.98\columnwidth]{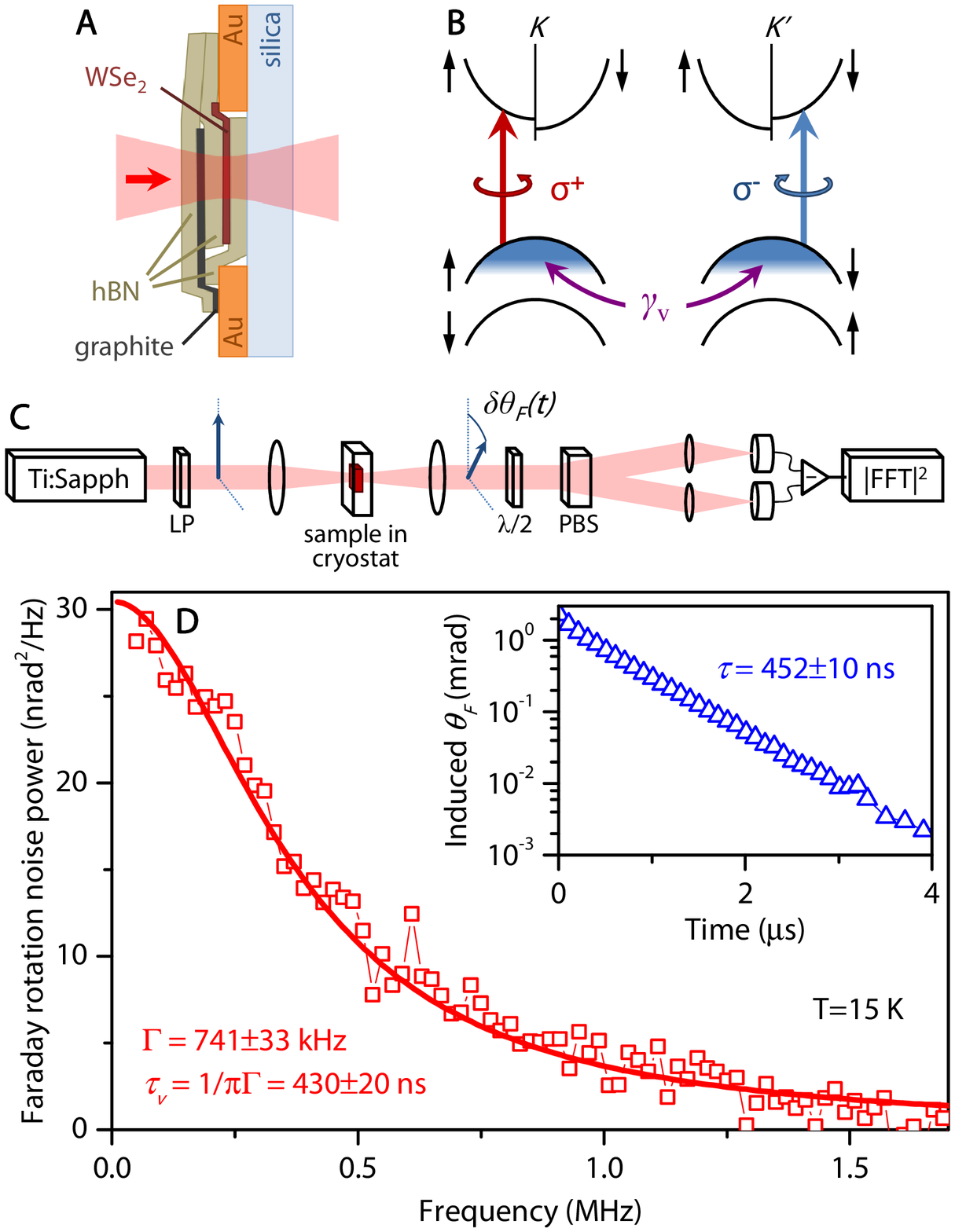}
\caption{\textbf{Sample, experimental setup, and valley noise spectrum of resident holes in monolayer WSe$_2$.} (\textbf{A}) Sample: A single WSe$_2$ monolayer is sandwiched between hBN layers and electrically gated. (\textbf{B}) Band structure and $\sigma^\pm$ optical transitions of hole-doped monolayer WSe$_2$. Even in thermal equilibrium, resident holes spontaneously scatter between $K$ and $K'$ valleys, giving a randomly fluctuating valley polarization noise. (\textbf{C}) To detect valley noise, a CW probe laser is linearly polarized and focused through the sample.  Thermodynamic valley fluctuations impart Faraday rotation fluctuations $\delta \theta_F (t)$ on the probe laser, which are detected using balanced photodiodes. (\textbf{D}) The valley noise power spectrum (symbols) of resident holes in monolayer WSe$_2$. Its Lorentzian lineshape (solid line) with full-width $\Gamma$ indicates an exponentially-decaying valley correlation with relaxation timescale $\tau_v$=$1/\pi \Gamma$=430$\pm$20~ns. Inset: valley relaxation measured separately in a perturbative pump-probe experiment.}
\label{setup}
\end{figure}

Figure \ref{setup}A depicts the sample, which is an exfoliated WSe$_2$ monolayer sandwiched between 25~nm thick slabs of hexagonal boron nitride (hBN). To electrostatically gate the resident carrier density in the WSe$_2$, part of the monolayer contacts a gold pad, and a separately-contacted thin graphite layer serves as a top gate. The sample is assembled on a transparent silica substrate, and covered with a final hBN layer for mechanical stability. Unless otherwise stated, all measurements were performed in the lightly hole-doped regime using a fixed gate voltage $V_g$ = +2~V (giving an estimated hole density of $1.2 \times 10^{12}$/cm$^2$). All results presented here were also qualitatively confirmed on a second sample, with nominally the same structure.

Figure \ref{setup}B shows a simple band diagram of a hole-doped WSe$_2$ monolayer, along with the optical selection rules in the $K$ and $K'$ valleys that couple to right- and left-circularly polarized ($\sigma^+$ and $\sigma^-$) light, respectively. The absorption of $\sigma^\pm$ light depends sensitively on the densities of resident holes, $p^\pm$ (especially at energies $E$ near the positively-charged exciton resonance, as discussed in detail later).  Associated with the $\sigma^\pm$ absorptions are the dispersive indices of refraction, $n^\pm (E)$. The \textit{difference} between these indices can be measured by optical Faraday rotation (FR), $\theta_F$. By definition, $\theta_F (E) \propto n^+ (E) - n^- (E)$, which in turn is inherently sensitive to the holes' valley polarization, $p^+ - p^-$. For these reasons, $\theta_F$ (and its close relative, Kerr rotation $\theta_K$) is often used to probe the nonequilibrium valley polarizations that are induced in pump-probe optical studies 
\cite{Yang_2015_NaturePhys, Song_2016_NanoLett,Dey_2017_PRL, Yan_2017_PRB_WSe2Relax2ns, Kim_2017_ScienceAdv_WSe2MoS2Relax40us,Volmer_2017_PRB,McCormick_2018_2Dmat}. 

In thermal equilibrium (and in zero magnetic field), the time-averaged valley polarization of the hole Fermi sea is strictly zero: $\langle p^+ - p^- \rangle=0$, $\langle n^+ - n^- \rangle=0$, and therefore $\langle \theta_F(t) \rangle=0$. However, thermodynamic fluctuations always exist, and holes in the $K$ valley can spontaneously scatter to the $K'$ valley (or vice-versa) with intrinsic rate $\gamma_v = 1/\tau_v$, leading to a small valley polarization that fluctuates randomly in time about zero. This ``valley noise'' can in principle be detected as a fluctuating Faraday rotation $\delta \theta_F(t)$.  The temporal correlation function of this valley noise, $S(t) = \langle \delta \theta_F(0) \delta \theta_F(t) \rangle$, will be determined by the intrinsic timescales of valley scattering and relaxation in the TMD monolayer \cite{Kubo_1966_RPP}. This overall approach is analogous to studies of optical spin noise spectroscopy in atomic alkali vapors \cite{Crooker_2004_Nature,Zapasskii_2013_AdvOptPhoton} and certain conventional semiconductors \cite{Oestreich_2005_PRL,Crooker_2009_PRB,Zapasskii_2013_PRL} in which $\sigma^\pm$ light couples to specific spin (but not valley) states due to the presence of spin-orbit coupling. 

The experiment is shown in Fig. \ref{setup}C. The sample is mounted on the cold finger of a small optical cryostat. A continuous-wave (CW) probe laser is linearly polarized and focused to a small (3-10~$\mu$m diameter) spot on the sample. We typically tune the laser well \textit{below} the charged exciton ($X^+$) transition energy and use small (tens of $\mu$W) power, to avoid any interband excitation of the monolayer. Stochastic valley noise imparts small FR fluctuations $\delta \theta_F(t)$ on the transmitted probe laser, which are detected with sensitive balanced photodiodes.  This fluctuating signal is amplified, digitized, and its power spectrum is computed and signal-averaged in real time using fast Fourier transform methods. 

Figure \ref{setup}D shows the main result, which is the direct observation of thermodynamic valley fluctuations from resident carriers (holes) in a single monolayer of the archetypal TMD  semiconductor WSe$_2$. The probe laser is tuned to 1.694~eV, which is $\sim$10~meV below the $X^+$ resonance and therefore in the forbidden bandgap of the WSe$_2$. The detected valley noise is very small, amounting to only a few tens of nrad$^2$/Hz of FR noise power, which is typically $\sim$0.1\% of the background (uncorrelated) noise power that is due to fundamental photon shot noise and amplifier noise. To subtract off this constant background noise, what Fig. \ref{setup}D shows is actually the \textit{difference} of two noise spectra: one acquired on the WSe$_2$ flake, and one acquired just off the flake, where the probe beam passes only through the hBN layers and the substrate. 

The resulting noise spectrum is peaked at zero frequency and has a Lorentzian lineshape, with full-width at half-maximum $\Gamma=741\pm$33~kHz.  Because the power spectrum of $\delta \theta_F(t)$ is equivalent (per the Wiener-Khinchin theorem \cite{Kubo_1966_RPP}) to the Fourier transform of the temporal correlation function $S(t) = \langle \delta \theta_F(0) \delta \theta_F(t) \rangle$, a Lorentzian noise spectrum indicates that $S(t)$ decays exponentially, with a single time scale. This time scale is precisely the intrinsic valley relaxation time $\tau_v$, which is related to the inverse-width of the noise power spectrum: $\tau_v = 1/(\pi \Gamma)$. Therefore, the noise spectrum in Fig. \ref{setup}D reveals a long valley relaxation time $\tau_v=430\pm$20~ns for resident holes in this WSe$_2$ monolayer (at 15~K and at this hole density). In this way, the intrinsic timescales of valley relaxation in monolayer TMD semiconductors are revealed simply by passively ``listening'' to the thermodynamic valley fluctuations alone, and without ever perturbing, pumping, or exciting the Fermi sea of resident holes away from equilibrium. 

Such a long intrinsic $\tau_v$ is consistent with very recent reports of extremely long valley relaxation of resident holes in WSe$_2$ \cite{Kim_2017_ScienceAdv_WSe2MoS2Relax40us,Dey_2017_PRL}, which result from strong spin-valley locking in the valence bands of monolayer TMDs \cite{Xiao_2012_PRL}. As depicted in Fig. \ref{setup}B, for a hole to scatter/relax between the upper valence bands in the $K$ and $K'$ valleys it must not only significantly change its momentum, but must \textit{also} simultaneously flip its spin. The low probability for this to occur means that measured relaxation timescales can be long, of order $\mu$s at low temperatures.

We use the noise-based methodology to validate conventional (\textit{i.e.}, perturbative) pump-probe methods, to resolve concerns about the role of dark excitons (or other experimentally-related perturbations) in studies of slow valley relaxation. We directly compare $\tau_v$ obtained from the valley noise to the decay time measured independently by optical pump-probe experiments. Using pulsed lasers, we perform a time-resolved Faraday rotation (TRFR) study of the same WSe$_2$ monolayer (at the same $T$ and $V_g$), using methods described previously \cite{Dey_2017_PRL}.  Here, 1.922~eV pump pulses inject valley-polarized electrons and holes into the WSe$_2$, which scatter, relax, and recombine, eventually perturbing the hole Fermi sea and generating a nonequilibrium valley polarization. The dissipative relaxation of the hole polarization back to equilibrium is monitored by the FR imparted on time-delayed 1.694~eV probe pulses. The nearly monoexponential decay of the induced valley polarization is shown in the inset of Fig. \ref{setup}D. The decay time is 450$\pm$10~ns, which is in close agreement with the valley correlation (relaxation) time $\tau_v$ inferred from passive detection of the equilibrium fluctuations alone. It is worth noting that, in contrast to many previous experiments \cite{Yang_2015_NaturePhys,Song_2016_NanoLett,Yan_2017_PRB_WSe2Relax2ns,Volmer_2017_PRB,McCormick_2018_2Dmat,Kim_2017_ScienceAdv_WSe2MoS2Relax40us,Dey_2017_PRL}, the decay does not show any fast ($<$10~ns) components, likely due to the high sample quality and hBN encapsulation. Because the valley noise detection is entirely passive and does not involve any optical pumping or excitation (as verified in more detail below), $\tau_v$ clearly cannot be due to the presence of optically-inactive ``dark'' excitons or trions.  Importantly, the very close agreement between the valley decay time measured by conventional TRFR and $\tau_v$ measured from valley noise therefore validates that long-lived TRFR signals, observed here and in many previous studies, \textit{also} arise principally from the valley polarization of resident carriers, and do not arise from dark states.

\begin{figure} [tbp]
\center
\includegraphics[width=.97\columnwidth]{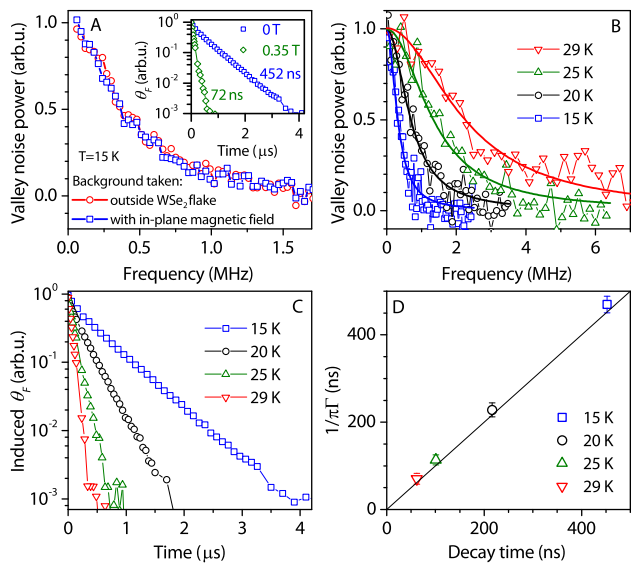}
\caption{\textbf{Comparing the valley noise measurements with perturbative pump-probe experiment.} (\textbf{A}) Valley noise spectra acquired using two different background subtraction methods: on/off the WSe$_2$ (red), and absence/presence of in-plane magnetic field $B_x$ (blue). Inset: The valley relaxation measured independently by TRFR at $B_x=0$ and 350~mT.  (\textbf{B}) Valley noise spectra of hole-doped WSe$_2$ at different temperatures (solid lines are Lorentzian fits). (\textbf{C}) Hole valley relaxation measured separately by TRFR. (\textbf{D}) Comparing $\tau_v$ (= $1/\pi \Gamma$) from the valley noise with the valley decay time measured by TRFR. 
}\label{correspondence}
\end{figure}

In Fig. \ref{correspondence} we vary the sample temperature and again compare the valley relaxation time $\tau_v$ deduced from noise experiments with the decay measured by conventional TRFR methods. Here we use a different means of background subtraction to isolate the valley noise: Instead of measuring noise spectra on and off the WSe$_2$ monolayer, we measure the noise in the absence and presence of an in-plane magnetic field $B_x$. TRFR studies of this sample, shown in the inset of Fig. \ref{correspondence}A, reveal that $B_x=350$~mT suppresses the measured valley decay time by nearly an order of magnitude.  We note that some earlier pump-probe studies of hole-doped WSe$_2$ showed very little dependence on $B_x$ \cite{Song_2016_NanoLett, Dey_2017_PRL}; however those samples were not hBN encapsulated and exhibited nonexponential decays, which may be related to their markedly different field dependence. Regardless, the observed reduction of the valley decay time means that the width $\Gamma$ of the associated valley noise spectrum increases (and its amplitude decreases) also by almost an order of magnitude, so that the spectrum at large $B_x$ is comparatively small and spectrally flat at low frequencies. It can therefore be used as a reliable background against which to compare measurements at $B_x=0$ (experimentally, it is more convenient to vary $B_x$ than reposition the probe laser on and off the sample).  To confirm that this background subtraction method works, Fig. \ref{correspondence}A shows the valley noise at 15~K obtained by both schemes; the agreement is very good. Fig. \ref{correspondence}B shows normalized noise spectra obtained at $T$ = 15, 20, 25, and 29~K.  All are well described by Lorentzian line shapes, with $\Gamma$ (= $1/\pi \tau_v$) increasing significantly to 4.5~MHz at 29~K.  Separately, conventional TRFR measurements of valley dynamics in this sample (Fig. \ref{correspondence}C) show strongly temperature-dependent decays. Fig. \ref{correspondence}D compares $\tau_v$ as measured by the two methods, again showing very close agreement and further validating the interpretation of long-lived TRFR signals as arising from carrier valley polarization. 

\begin{figure} [tbp]
\center
\includegraphics[width=0.9\columnwidth]{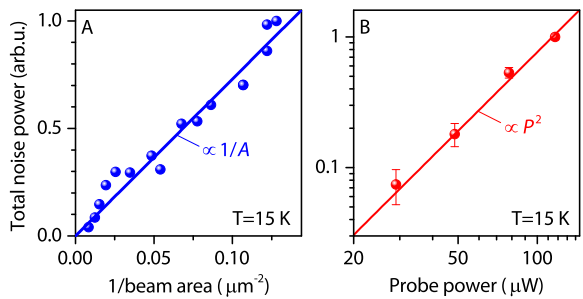}
\caption{\textbf{Dependence of the measured noise on probe parameters.} (\textbf{A}) The total (frequency-integrated) valley noise power scales as the \textit{inverse} of the cross-sectional area $A$ of the probe laser spot on the sample, as expected for optically-detected noise measurements. (\textbf{B}) The total valley noise scales quadratically with the power $P$ of the probe laser, again as expected for noise studies.}
\label{area}
\end{figure}

In order to confirm that the probe laser in the noise experiments does indeed behave as a passive detector of intrinsic valley fluctuations (and is not inadvertently perturbing the sample by, for example, exciting real carriers), we investigate the dependence of the measured noise on the parameters of the probe laser. Fig. \ref{area}A shows that the total frequency-integrated noise power scales \textit{inversely} with the cross-sectional area of the probe laser spot on the sample, which we vary by moving the sample in an out of the beam focus. An inverse-area dependence is a well-established hallmark of optically-detected noise signals, established from earlier studies of spin noise in atomic vapors \cite{Crooker_2004_Nature} and semiconductors \cite{Crooker_2009_PRB}. We briefly summarize the reason for this dependence as follows: Consider the ensemble of $N = pA$ holes within the probe laser spot, where $p$ is the hole density and $A$ is the spot area. The number of holes that randomly fluctuate between the $K$ and $K'$ valleys is proportional to $\sqrt{N}$. Therefore, the fluctuations of holes' valley polarization (which is probed by FR) scale with $\sqrt{N}/N = 1/\sqrt{N} = 1/\sqrt{pA}$, which \textit{increases} when $A$ shrinks and $N$ becomes small. At the smallest spot size, we estimate that the measured noise arises from valley fluctuations of order of $\pm$50 holes (assuming that only holes within $k_B T$ of the Fermi energy can scatter). Separately, Fig. \ref{area}B shows the dependence of the measured noise on the probe laser power. For a given fluctuation in the sample, doubling the probe power should trivially double the fluctuating voltage generated by the photodiodes, quadrupling the measured noise power (in units of volts-squared of total detected noise), which is indeed observed. Any deviations from the inverse-area dependence or the quadratic power dependence could indicate that the probe itself is pumping or perturbing the sample.  

\begin{figure} [tbp]
\center
\includegraphics[width=.98\columnwidth]{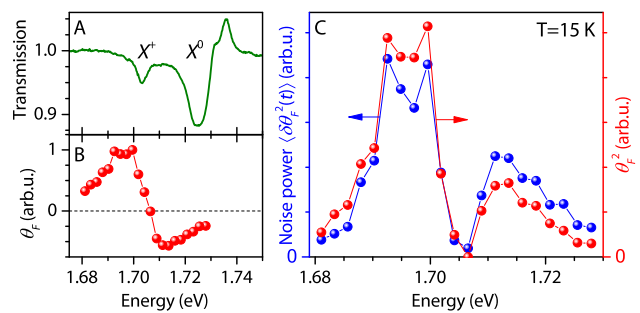}
\caption{\textbf{Dependence of the hole valley noise on the probe photon energy.} (\textbf{A}) Optical transmission spectrum of the lightly hole-doped WSe$_2$ monolayer ($V_g$ = +2~V), showing neutral ($X^0$) and positively-charged exciton ($X^+$) resonances. (\textbf{B}) Energy-dependent FR spectrum $\theta_F (E)$ induced by an \textit{intentional} (optically-pumped) valley polarization of the resident holes (see text). $\theta_F (E)$ is antisymmetric and centered on $X^+$, as expected. (\textbf{C}) Spectral dependence of the total valley noise power, $\langle \delta \theta^2_F(t) \rangle$ (blue points). The red points show the \textit{square} of $\theta_F (E)$ from panel (\textbf{B}), showing close agreement.}
\label{wavelength}
\end{figure}

We also investigate the dependence of the hole valley noise on the probe photon energy $E$, and show that the noise is largest when the probe laser is tuned near to -- but not on -- the $X^+$ charged exciton transition. Figure \ref{wavelength}A shows the transmission spectrum of the lightly hole-doped WSe$_2$ monolayer. The $X^+$ absorption appears as a very narrow (4.7~meV wide) resonance at 1.703~eV, which is $\sim$22~meV below the neutral exciton resonance ($X^0$ is still visible at this small hole density). Figure \ref{wavelength}B shows the result of a conventional CW pump-probe Faraday rotation experiment, in which a weak circularly-polarized CW pump laser (at 1.96~eV) generates a steady-state nonequilibrium valley polarization in the hole Fermi sea, while the induced FR spectrum $\theta_F (E)$ is concurrently measured by a tunable CW probe laser. Note that $\theta_F (E)$ shows a clear antisymmetric lineshape, in accordance with its dependence on the dispersive indices of refraction, $n^+(E) - n^-(E)$. Most importantly, the induced $\theta_F (E)$ is centered about the $X^+$ absorption resonance, in agreement with the expectation that charged exciton transitions (unlike neutral excitons) depend \textit{explicitly} on the presence of resident carriers -- \textit{i.e.}, by definition, the $\sigma^+ / \sigma^-$ polarized $X^+$ transition can only have oscillator strength if there are pre-existing (resident) holes in the $K' / K$ valleys, respectively.  

As such, we may therefore anticipate that thermodynamic valley fluctuations of the hole Fermi sea will generate a similar spectral response, and indeed we find this to be the case. The blue points in Fig. \ref{wavelength}C show the total valley noise power $\langle \delta \theta^2_F(t) \rangle$ versus the probe laser's energy. For comparison, red points show the \textit{square} of the induced $\theta_F (E)$ shown in Fig. \ref{wavelength}B, revealing a nearly identical spectral dependence. We emphasize that valley fluctuations can therefore be detected by FR using light tuned in energy well \textit{below} the lowest absorption resonance in the WSe$_2$ monolayer, which further assures the non-perturbative nature of this noise-based approach, in which valley dynamics can be studied quantitatively under conditions of strict thermal equilibrium. 

\begin{figure} [tbp]
\center
\includegraphics[width=.90\columnwidth]{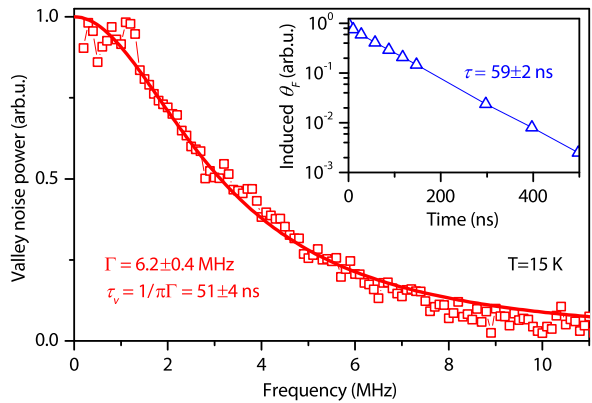}
\caption{\textbf{The valley noise spectrum of electron-doped monolayer WSe$_2$.} The spectrum shows Lorentzian lineshape and much larger width $\Gamma$ (indicating shorter valley relaxation time $\tau_v$) as compared to the hole-doped regime (\textit{cf.} Fig. \ref{setup}D). The spectrum was measured at $V_g = -2$~V and $T=15$~K. Inset: electron valley relaxation measured independently by conventional  (perturbative) TRFR, which corroborates the shorter $\tau_v$.}
\label{electron}
\end{figure}

Finally, Fig. \ref{electron} shows that valley noise is also detectable in a Fermi sea of resident \textit{electrons} in monolayer TMDs. Here, $V_g = -2$~V, giving a small electron density $n \approx 1.2 \times 10^{12}$/cm$^2$. The probe laser is tuned to 1.686 eV, or 7 meV below the negatively-charged exciton ($X^-$) absorption. The measured noise spectrum is again Lorentzian, but its width $\Gamma=6.2$~MHz  is nearly an order of magnitude larger than the width of the hole noise spectrum in this same WSe$_2$ monolayer (\textit{cf.} Fig. \ref{setup}D). This indicates a much faster intrinsic relaxation time ($1/\pi\Gamma=51$~ns) for electrons as compared to holes, in agreement with the expectation that spin-valley locking is much weaker in the conduction bands than in the valence bands of monolayer TMDs \cite{Dey_2017_PRL}. The inset of Fig. \ref{electron} shows a conventional TRFR study of this electron-doped monolayer, which closely corroborates the much faster relaxation of resident electrons, again confirming that (at least here) TRFR measurements are primarily sensitive to the valley polarization of resident electrons.  

These noise-based studies therefore demonstrate a promising new alternative approach for probing the valley dynamics of electrons and holes in TMD monolayers. Spontaneous fluctuations of the valley polarization in strict thermal equilibrium are shown to encode the intrinsic timescales of valley relaxation, and can be observed via sensitive Faraday rotation spectroscopy without any external pumping or excitation. These proof-of-concept measurements reveal the long, monoexponential \textit{intrinsic} valley relaxation of holes and electrons in hBN-encapsulated WSe$_2$ monolayers. Moreover, this methodology validates traditional (perturbative) time-resolved and CW pump-probe spectroscopies that show long decays in TMD monolayers doped with resident electrons and holes.  

The fluctuation-based methods demonstrated here should be particularly suitable for future high-quality TMD monolayers and heterostructures, for which the valley relaxation times are expected to be exceptionally long, such that features in the noise power spectra are narrow and large. We expect that under such conditions the \textit{intrinsic} valley relaxation timescales revealed by noise can be longer than the timescales accessible with perturbative methods, since any experimentally-related perturbation (such as dark excitons or trapped states) may be stronger than the very weak interactions that lead to slow relaxation. This opens new opportunities for studies of valley physics in novel 2D semiconductors.

We thank \L{}ukasz K\l{}opotowski for helpful discussions,  Andrew Balk for assistance with the digitizer, and we gratefully acknowledge support from the Los Alamos LDRD program.  Work at the NHMFL was supported by NSF DMR-1644779, the State of Florida, and the US DOE. Work at the University of Washington was supported by the DOE Basic Energy Sciences, Materials Sciences and Engineering Division (DE-SC0018171).

\end{document}